\documentclass[twocolumn,superscriptaddress,pra]{revtex4}
\usepackage{amsmath}
\usepackage{amssymb}
\usepackage{graphicx}
\usepackage{bm}

\begin{document}

\title{Quantum Fluctuations in Noether's Theorem}

\author{J. A. Crosse}
\email{jac32@nyu.edu}
\affiliation{New York University Shanghai, 567 West Yangsi Road, Shanghai, 200124, China.}
\affiliation{NYU-ECNU Institute of Physics at NYU Shanghai, 3663 Zhongshan Road North, Shanghai, 200062, China.}

\date{\today}

\begin{abstract}
When deriving the expression for the Noether's current one obtains a term that is an integral over the Euler-Lagrange equation. This term vanishes along the classical path and so does not contribute to the current. However, quantum particles do not necessarily follow the classical path; this concept is the basis of Feynman's path integral quantization formalism. Thus, for a quantum system, the Euler-Lagrange term does not vanish and, hence, corresponds to a quantum fluctuation of the Noether's current owing to the system deviating from the classical path. Here, we compute this quantum fluctuation from Noether's theorem and the path integral formalism. We find that, for a time invariant system, the expression for the quantum fluctuations is give by minus the functional expression for the Hamiltonian, $\Delta E = -H(y)$, evaluated at the distance from the classical path, $y$. As an example we evaluate quantum energy fluctuations of a single Harmonic oscillator. In the short time limit we recover the energy-time uncertainty relation and in the long time limit we recover the zero point energy of the quantum harmonic oscillator. 
\end{abstract} 

\maketitle

\section{Introduction}

The action, $S$, is the fundamental mathematical expression that describes any physical system. From the principle of stationary action one can find the evolution of the system via the Euler-Lagrange equations \cite{EL1,EL2} and the constants of motion can be found from the continuous symmetries of the action via Noether's theorem \cite{noether}. 

In the derivation of the conserved Noether's currents one finds a contribution that is an integral over the Euler-Lagrange equation. This term is usually set to zero because the Euler-Lagrange equation vanishes along the classical path. At this point the system is fixed to being classical; by enforcing the validity of the Euler-Lagrange equation the system is constrained to follow the classical path. However, quantum systems do not necessarily follow the classical path. In fact, Feynman's path integral quantization formalism specifically sums the weighted contribution from all possible paths \cite{path}. In this case, the Euler-Lagrange term would not vanish and would contribute to the conserved current. This contribution is the result of the system deviating from the classical path and, hence, corresponds to a quantum fluctuation of the conserved current. 

Here, we calculate the quantum fluctuation contribution to the Noether's current \cite{sup}. We achieve this by relaxing the constraint that the system must follow the classical path and then computing the contribution to the Euler-Lagrange term from all possible paths. We find that, starting with only the classical action and only using the concept that the quantum evolution is a weighted sum of all possible paths, we are able to recover the key quantum properties of the equivalent quantized system. 

\section{Noether's Theorem in Classical Mechanics}

Noether's Theorem \cite{noether} states that for every continuous symmetry of the action, $S$, there is an associated conserved current. This is usually shown by varying the action and setting it equal to a boundary term
\begin{equation}
\delta S = \int_{t_i}^{t_f}dt\,\left[\frac{\partial  L}{\partial x}\delta x + \frac{\partial  L}{\partial \dot{x}}\delta \dot{x}\right] = \Lambda(t)\bigg|_{t_i}^{t_f},
\end{equation}
where the Lagrangian, $L=K(\dot{x})-U(x)$, is the difference between the kinetic, $K$, and potential, $U$, energies. With some rearrangement one finds
\begin{multline}
\int_{t_i}^{t_f}dt\,\left\{\frac{d}{dt}\left[\left(\frac{\partial  L}{\partial \dot{x}}\delta x\right)  - \Lambda\right] \right.\\
\left. + \left[\frac{\partial  L}{\partial x} - \frac{d}{dt}\left(\frac{\partial  L}{\partial \dot{x}}\right)\right]\delta x \right\} = 0.
\label{noether}
\end{multline}
The second term in Eq. \eqref{noether} vanishes via the Euler-Lagrange equation
\begin{equation}
\frac{\partial  L}{\partial x} - \frac{d}{dt}\left(\frac{\partial  L}{\partial \dot{x}}\right) = 0,
\end{equation}
which defines the classical path. For the overall integral to vanish the first term in Eq. \eqref{noether} must also vanish. Therefore
\begin{equation}
\frac{d}{dt}\left[\left(\frac{\partial  L}{\partial \dot{x}}\delta x\right)  - \Lambda\right] = 0,
\end{equation}
which implies that the expression in square brackets does not change with time and, hence, 
\begin{equation}
J_{cl} = \left(\frac{\partial  L}{\partial \dot{x}}\delta x\right)  - \Lambda,
\end{equation}
is a constant of motion. The quantity, $J_{cl}$, is known as the (classical) Noether's current.

\section{Noether's Theorem in Quantum Mechanics}

In Feynman's path integral formalism of quantum mechanics, the evolution of a quantum system is the weighted sum over all possible paths not just the classical one. Thus, in this description, the second term in Eq. \eqref{noether} does not necessarily vanish and we have an additional term in the Noether's current. If one writes 
\begin{equation}
\frac{dF}{dt} = \left[\frac{\partial  L}{\partial x} - \frac{d}{dt}\left(\frac{\partial  L}{\partial \dot{x}}\right)\right]\delta x,
\label{f}
\end{equation}
we then have
\begin{equation}
\int_{t_i}^{t_f}dt\,\left\{\frac{d}{dt}\left[\left(\frac{\partial  L}{\partial \dot{x}}\delta x\right)  - \Lambda\right]+\frac{dF}{dt}\right\} = 0.
\end{equation}
Now, if the overall integral is to vanish, one requires
\begin{equation}
\frac{d}{dt}\left[\left(\frac{\partial  L}{\partial \dot{x}}\delta x\right)  - \Lambda + F\right]= 0,
\end{equation}
and, hence, the conserved current becomes
\begin{equation}
J = \left(\frac{\partial  L}{\partial \dot{x}}\delta x\right)  - \Lambda + F = J_{cl} + F.
\label{jq}
\end{equation}
Along the classical path $F=0$. Off the classical path this is not necessarily the case. Thus, $F$ represents a quantum fluctuation caused by the system deviating from the classical path.

\section{Quantum Energy Fluctuations}

As an example one can consider an action that is invariant in time (i.e. $t' = t + \epsilon$, leaves the action unchanged). A short calculation leads to
\begin{equation}
\delta x = \dot{x} \qquad\mathrm{and}\qquad \frac{d\Lambda}{dt} = \frac{dL}{dt},
\end{equation}
and, hence, the conserved current is 
\begin{equation}
J = \left(\frac{\partial  L}{\partial \dot{x}}\dot{x}\right)  - L = H,
\end{equation}
the Hamiltonian or total energy.

To find an expression for the quantum fluctuations, $F$, one can expand the Lagrangian about the classical path $x\rightarrow x_{cl} + y$
\begin{widetext}
\begin{align}
L(x_{cl} + y,\dot{x}_{cl} + \dot{y}) &= L(x_{cl},\dot{x}_{cl}) + \left\{y\left.\frac{\partial L(x,\dot{x})}{\partial x}\right|_{x=x_{cl}}+\dot{y}\left.\frac{\partial L(x,\dot{x})}{\partial \dot{x}}\right|_{x=x_{cl}}\right\}\nonumber\\
&\hspace{2.5cm}+ \frac{1}{2}\left\{y^2\left.\frac{\partial^2 L(x,\dot{x})}{\partial x^2}\right|_{x=x_{cl}}+2y\dot{y}\left.\frac{\partial^2 L(x,\dot{x})}{\partial \dot{x}\partial x}\right|_{x=x_{cl}}+\dot{y}^2\left.\frac{\partial^2 L(x,\dot{x})}{\partial \dot{x}^2}\right|_{x=x_{cl}}\right\} + \cdots\nonumber\\
\label{ex}
\end{align}
\end{widetext}
The zeroth order term is the Lagrangian along the classical path and, hence, obeys the Euler-Lagrange equation. Thus, its contribution to the quantum fluctuations vanishes. The first order terms are the variation of the Lagrangian (action). As the classical path, by definition, extremizes the action these terms also vanishes. Hence, the second order terms are the lowest order terms that contribute to the quantum fluctuations.

If the Lagrangian is bilinear or quadratic in $x$ and $\dot{x}$ then the second order derivatives will result in two times the constant prefactor for that term. Thus, the expansion retains the functional form of the Lagrangian but replaces the variables $x$ and $\dot{x}$ in the with $y$ and $\dot{y}$, $L(x,\dot{x})\rightarrow L(y,\dot{y})$. Specifically for bilinear or quadratic Lagrangians, these terms will be the highest order term in the expansion (all other terms will vanish) and so the expansion is exact.

One can find an expression for the quantum fluctuations by changing variables $x\rightarrow x_{cl} + y$ (the classical path is fixed so $x_{cl}$ is constant) and evaluating Eq. \eqref{f} for the second order term in the expansion in Eq. \eqref{ex}
\begin{equation}
\frac{dF}{dt} = \left[\frac{\partial L(y,\dot{y})}{\partial y} - \frac{d}{dt}\left(\frac{\partial L(y,\dot{y})}{\partial \dot{y}}\right)\right]\dot{y},
\end{equation}
where we have used $\delta y = \dot{y}$, which is the variation required for a time symmetric Lagrangian and, hence, fluctuations in energy. One can simplify this expression by noting that the total derivative of the Lagrangian is
\begin{equation}
\frac{dL}{dt} = \frac{\partial L}{\partial y}\frac{dy}{dt}+\frac{\partial L}{\partial \dot{y}}\frac{d\dot{y}}{dt}.
\end{equation}
Hence, one finds
\begin{align}
\frac{dF}{dt} &= \left[\frac{\partial L(y,\dot{y})}{\partial y} - \frac{d}{dt}\left(\frac{\partial L(y,\dot{y})}{\partial \dot{y}}\right)\right]\frac{dy}{dt},\nonumber\\
&= \frac{d}{dt}\left[L(y,\dot{y}) - \frac{\partial L(y,\dot{y})}{\partial \dot{y}}\dot{y}\right],\nonumber\\
&= \frac{d}{dt}\left[-H(y,\dot{y})\right],
\end{align}
where $H(y,\dot{y})$ is functional expression for the Hamiltonian of the system but in terms of the variable $y$, the deviation from the classical path. From the definition in Eq. \eqref{jq} we can see that the quantum energy fluctuations are given by
\begin{equation}
F = -H(y,\dot{y}).
\end{equation}

\section{Quantum Energy Fluctuations of the Harmonic Oscillator}

\subsection{General Expression}

To see how this formalism works in a practice we can compute the quantum energy fluctuations for a specific system. The Lagrangian for the Harmonic oscillator with mass $m$ and angular frequency $\omega$ is given by
\begin{equation}
L(x,\dot{x}) = \frac{1}{2}m\dot{x}(t)^2 - \frac{1}{2}m\omega^2 x(t)^2.
\end{equation}
As a consistency check for the above description we can perform the functional expansion about the classical path for this specific Lagrangian
\begin{multline}
L(x_{cl} + y,\dot{x}_{cl} + \dot{y}) = L(x_{cl}) + \int dt'\,y(t')\left.\frac{\delta L(x)}{\delta x(t')}\right|_{x=x_{cl}} \\
+ \frac{1}{2}\int dt''\int dt'\,y(t'')y(t')\left.\frac{\delta^{2} L(x)}{\delta x(t'')\delta x(t')}\right|_{x=x_{cl}} + \cdots
\end{multline}
The first order term we would expect to vanish as this defines the classical path. Evaluating the first term in the expansion gives
\begin{align}
&L^{(1)}(y,\dot{y}) = \int dt'\,y(t')\left.\frac{\delta L(x)}{\delta x(t')}\right|_{x=x_{cl}} ,\nonumber\\
&= \int dt'\,y(t')\left. \frac{\delta}{\delta x(t')}\left[\frac{1}{2}m\dot{x}(t)^2 - \frac{1}{2}m\omega^2x(t)^2\right]\right|_{x=x_{cl}} ,\nonumber\\
&=  \int dt'\,y(t')\left.\left[m\dot{x}(t)\frac{d}{dt}\delta(t-t') - m\omega^2x(t)\delta(t-t')\right]\right|_{x=x_{cl}},\nonumber\\
&= -m \,y(t)\left[\ddot{x}_{cl}(t) + \omega^2x_{cl}(t)\right].
\end{align}
The term in square brackets is just the Euler-Lagrange equation of motion for the harmonic oscillator. As this is evaluated along the classical path this term vanishes identically. The second order term is computed as follows
\begin{widetext}
\begin{align}
L^{(2)}(y,\dot{y}) &= \frac{1}{2}\int dt''\int dt'\,y(t'')y(t')\left.\frac{\delta^{2} L[x]}{\delta x(t'')\delta x(t')}\right|_{x=x_{cl}}\nonumber\\
&=\frac{1}{2}\int dt''\int dt'\,y(t'')y(t')\left.\frac{\delta^{2}}{\delta x(t'')\delta x(t')}\left[\frac{1}{2}m\dot{x}(t)^2 - \frac{1}{2}m\omega^2x(t)^2\right]\right|_{x=x_{cl}}\nonumber\\
& = \frac{1}{2}\int dt''\int dt'\,y(t'')y(t')\left[m\frac{d}{dt}\delta(t-t'')\frac{d}{dt}\delta(t-t') - m\omega^2\delta(t-t'')\delta(t-t')\right],\nonumber\\
& = \frac{m}{2}\left[\dot{y}(t)^2 - \omega^2y(t)^2\right],
\end{align}
\end{widetext}
which is the usual second order expansion of the Lagrangian that is found when evaluating the Feynman path integral for the harmonic oscillator \cite{nakahara}. Substituting this back into Eq. \eqref{f} we find
\begin{align}
\frac{dF}{dt} &= \left[\frac{\partial  L^{(2)}(y,\dot{y})}{\partial y} - \frac{d}{dt}\left(\frac{\partial  L^{(2)}(y,\dot{y})}{\partial \dot{y}}\right)\right]\delta y ,\nonumber\\
&= \left[-m\ddot{y}(t) - m\omega^2y(t)\right]\dot{y},\nonumber\\
&= \frac{d}{dt}\left[-\frac{1}{2}m\dot{y}(t)^2 - \frac{1}{2}m\omega^2y(t)^2\right].
\end{align}
Thus, we find that the energy fluctuations of the harmonic oscillator are given by
\begin{equation}
F = -H(y,\dot{y}),
\end{equation}
as expected.

\subsection{Time Averaged Energy Fluctuations}

Using the expression for $F$ in terms of the Hamiltonian, $H$, one can compute an expression for the time averaged quantum energy fluctuations over a time interval, $\Delta t$, of a single Harmonic oscillator. In the path integral formalism the time averaged expectation value of, $F$, is given by \cite{pands}
\begin{equation}
\Delta E = \frac{1}{\Delta t}\int_{0}^{\Delta t}dt\,\langle F(t) \rangle = \frac{1}{\Delta t}\int_{0}^{\Delta t}dt\,\left\{\frac{\displaystyle\int\mathcal{D}y\, F\, e^{iS/\hbar}}{\displaystyle\int\mathcal{D}y\,e^{iS/\hbar}}\right\}.
\end{equation}
Substituting in the expression for $F$ and using the expansion of the action, $S$, about the classical path (as is the standard method for evaluating the Harmonic Oscillator path integral \cite{nakahara}) and noting that the denominator is independent of $t$ (it is only a function of $\Delta t$), one finds
\begin{widetext}
\begin{equation}
\Delta E = \frac{e^{iS_{cl}/\hbar}\displaystyle\int\mathcal{D}y\,\left\{\frac{1}{\Delta t}\int_{0}^{\Delta t}dt\,\left(-\frac{m}{2}\right)\left[\dot{y}(t)^2 + \omega^2y(t)^2\right]\right\} e^{\displaystyle i\frac{m}{2\hbar}\int_{0}^{\Delta t}dt\,\left[\dot{y}(t)^2 - \omega^2y(t)^2\right]}}{\displaystyle\int\mathcal{D}y\,e^{iS/\hbar}},
\end{equation}
\end{widetext}
where $S_{cl}$ is the action along the classical path. We can evaluate the path integral in the usual way by substituting in the Fourier series expansion for $y$ \cite{nakahara}
\begin{gather}
y = \sum_{j=1}^{N}a_n\sin\left(\frac{j\pi t}{\Delta t}\right),\\
\dot{y} = \sum_{j=1}^{N}a_n\left(\frac{j\pi}{\Delta t}\right)\cos\left(\frac{j\pi t}{\Delta t}\right),
\end{gather}
which we write as a finite series of $N$ terms but will eventually take $N\rightarrow\infty$. After evaluating the resulting Gaussian integrals one obtains
\begin{equation}
\Delta E = \lim_{N\rightarrow \infty}\frac{i\hbar}{2\Delta t}\sum_{j=1}^{N-1}\frac{\displaystyle\left[\left(\frac{j\pi}{\Delta t}\right)^2 + \omega^2\right]}{\displaystyle\left[\left(\frac{j\pi}{\Delta t}\right)^2 - \omega^2\right]}.
\end{equation}
The sum can be split into two parts by adding and subtracting a factor of $\omega^2$ to the numerator
\begin{equation}
\Delta E = \lim_{N\rightarrow \infty}\frac{i\hbar}{2\Delta t}\left\{\sum_{j=1}^{N-1}1 + \sum_{j=1}^{N-1}\frac{\displaystyle 2\omega^2}{\displaystyle\left[\left(\frac{j\pi}{\Delta t}\right)^2 - \omega^2\right]}\right\}.
\label{dE}
\end{equation}
The first sum easily evaluates to
\begin{equation}
\sum_{j=1}^{N-1}1 = N-1.
\end{equation}
part of which diverges when the limit as $N\rightarrow\infty$ is taken. This corresponds to an infinite vacuum energy.  The second sum evaluates to a rather complicated expression involving the digamma function, $\psi^{(0)}(z)$,
\begin{align}
&\sum_{j=1}^{N-1}\frac{\displaystyle 2\omega^2}{\displaystyle\left[\left(\frac{j\pi}{\Delta t}\right)^2 - \omega^2\right]} = 1-\omega\Delta t\cot\left(\omega\Delta t\right) \nonumber\\
&\hspace{1cm}+ \frac{\omega\Delta t}{\pi}\left[\psi^{(0)}\left(N - \frac{\omega\Delta t}{\pi}\right) - \psi^{(0)}\left(N + \frac{\omega\Delta t}{\pi}\right)\right].
\end{align}
However, in the limit as $N\rightarrow\infty$ the digamma functions cancel. Thus, finite part of the average vacuum energy fluctuations becomes
\begin{equation}
\Delta E = -\frac{i\hbar\omega}{2}\cot\left(\omega\Delta t\right). 
\label{E}
\end{equation}
Note that the factor of `$1$' from the second summation is canceled by the factor of `$-1$' from the first summation. The form of Eq. \eqref{E} is reminiscent of the expressions for the vacuum energy in studies of the Casimir force between two perfectly conducting plates \cite{milton}, except with the wavenumber replaced by the angular frequency and the plate separation replaced with the time interval.

If one expands the expression in Eq. \eqref{E} for small $\Delta t$ one obtains
\begin{equation}
\Delta E \approx -\frac{i\hbar}{2\Delta t} + \frac{i\hbar}{6}\omega^2\Delta t + \mathcal{O}\left(\Delta t^3\right),
\end{equation}
which to leading order recovers the energy-time uncertainty relation up to an irrelevant complex phase
\begin{equation}
\Delta E \Delta t \approx -i\frac{\hbar}{2}.
\end{equation}
We can find the average vacuum energy fluctuations in the long time limit by adding a small imaginary element then taking the time interval to infinity, $\Delta t \rightarrow \infty(1-i\epsilon)$. This techniques is often used in quantum field theory to project out the vacuum states \cite{pands}. On evaluating this limit one finds
\begin{equation}
\Delta E = \lim_{\Delta t \rightarrow \infty(1-i\epsilon)}-\frac{i\hbar\omega}{2}\cot\left(\omega\Delta t\right) = \frac{\hbar\omega}{2},
\end{equation}
which is the zero-point energy for a single Harmonic oscillator.

\section{Summary}

Here we have shown that Noether's theorem includes quantum fluctuations if one allows the evolution of the system to deviate from the classical path, an idea which itself is the conceptual basis for Feynman's path integral quantization formalism. Hence, if the classical action is known, it is possible to obtain the quantum fluctuations of any conserved quantity directly from Noether's theorem. For a single quantum harmonic oscillator we recover a familiar expression that appear in studies the quantum fluctuations that drive the Casimir force. In the short time limit we recover the energy-time uncertainty relation and in the long time limit we obtain the zero point energy of a single quantum Harmonic oscillator. 

It is worth emphasizing that the action used here is the time integral over the classical Lagrangian, i.e. the Lagrangian that gives the classical equation of motion for the the classical harmonic oscillator, not the Lagrangian that gives the quantum (Schr\"odinger) equation of motion for the quantum Harmonic oscillator. The Lagrangian that gives the equation of motion for the quantum Harmonic oscillator would be 
\begin{equation}
L= \frac{i\hbar}{2}\left(\psi^{\ast}\frac{\partial\psi}{\partial t} - \frac{\partial\psi^{\ast}}{\partial t}\psi\right)-\frac{\hbar^2}{2m}\frac{\partial \psi^{\ast}}{\partial x}\frac{\partial \psi}{\partial x}-\frac{1}{2}m\omega^2x^2\psi^{\ast}\psi.
\end{equation}
varied with respect to $\psi^{\ast}$, which would give a different result and, hence, the usual quantum energy properties would not be recovered. This observation may have relevance in quantum field theory where for particle fields one usually starts with a Lagrangian that recovers the first quantized Klein-Gordon or Dirac equations rather than a fully classical counterpart.



\begin{thebibliography}{99}

\bibitem{EL1}
L. Euler, Elementa Calculi Variationum, (1766) (in Latin).

\bibitem{EL2}
J.-L. Lagrange, Mécanique Analytique, (1788) (in French).

\bibitem{noether}
E. Noether, Goett. Nachr \textbf{1918}, 235 (1918) (in German).

\bibitem{path}
R. P. Feynman, Rev. Mod. Phys. \textbf{20}, 367 (1948).

\bibitem{sup}
See Supplementary Information.

\bibitem{nakahara}
M. Nakahara, \textit{Geometry, Topology and Physics, 2nd Ed.}, (Institute of Physics Publishing, London, 2003). 

\bibitem{pands}
M. E. Peskin and D. V. Schroeder, \textit{An Introduction to Quantum Field Theory}, (Perseus Books publishing, Massachusetts, 1995). 

\bibitem{milton}
K. Milton, \textit{The Casimir Effect : Physical Manifestations Of Zero-point Energy}, (World Scientific Publishing, Singapore, 2001). 




\end{thebibliography}
\end{document}